\documentclass{vldb}
\usepackage{graphicx}
\usepackage{enumitem}

\usepackage{pgfplots}
\pgfplotsset{compat=1.14}

\usepackage{balance}  
\usepackage{array}

\usepackage[hyphens]{url}

\vldbTitle{eXtreme Modelling in Practice}
\vldbAuthors{A. Jesse Jiryu Davis, Max Hirschhorn, Judah Schvimer}
\vldbDOI{https://doi.org/10.14778/3397230.3397233}
\vldbVolume{13}
\vldbNumber{9}
\vldbYear{2020}

\begin{document}

\emergencystretch 3em


\title{eXtreme Modelling in Practice}



%
%
%
%

\numberofauthors{3} %
\author{
%
\alignauthor
A. Jesse Jiryu Davis\\
       \affaddr{MongoDB, Inc.}\\
       \affaddr{1633 Broadway}\\
       \affaddr{New York, NY 10019}\\
       \email{jesse@mongodb.com}
\alignauthor
Max Hirschhorn\\
       \affaddr{MongoDB, Inc.}\\
       \affaddr{1633 Broadway}\\
       \affaddr{New York, NY 10019}\\
       \email{max.hirschhorn @mongodb.com}
\alignauthor
Judah Schvimer\\
       \affaddr{MongoDB, Inc.}\\
       \affaddr{1633 Broadway}\\
       \affaddr{New York, NY 10019}\\
       \email{judah@mongodb.com}
}


\maketitle

\pagestyle{empty}
\setcounter{page}{1346}

\begin{abstract}
Formal modelling is a powerful tool for developing complex systems. 
At MongoDB, we use TLA\textsuperscript{+} to model and verify multiple aspects of several systems. 
Ensuring conformance between a specification and its implementation can add value to any specification; it can avoid transcription errors, prevent bugs as a large organization rapidly develops the specified code, and even keep multiple implementations of the same specification in sync.
In this paper, we explore model-based testing as a tool for ensuring specification-implementation conformance.
We attempted two case studies: model-based trace-checking (MBTC) in the MongoDB Server's replication protocol and model-based test-case generation (MBTCG) in MongoDB Realm Sync's operational transformation algorithm.
We found MBTC to be impractical for testing that the Server conformed to a highly abstract specification. 
MBTCG was highly successful for Realm Sync, however.
We analyze why one technique succeeded and the other failed, and advise future implementers making similar attempts at model-based testing.
\end{abstract}

\section{Introduction}
\label{sec:introduction}

It is well established that formal modelling catches bugs, aids software design, and improves code quality \cite{Newcombe2014UseOfFormalMethodsAmazon}.
Formal methods are used increasingly to gain confidence in mission-critical, highly complex systems in multiple domains across many companies. 
Amazon \cite{Chudnov18AmazonS2N, Cook18SecurityAWS, Newcombe2014UseOfFormalMethodsAmazon}, Intel \cite{Beers08IntelExperience, Kaivola09IntelI7}, Microsoft \cite{Shukla18AzureCosmosDB}, and Springer Verlag \cite{Neubauer12AutomatedContinuousQualityAssurance} have written about their uses of formal methods and the value they've gained from modelling and verifying industrial software. 
But software engineers in industry are often skeptical of a model's value unless it is shown to match the implementation \cite{Newcombe2014UseOfFormalMethodsAmazon, Wayne18AgileFormalMethods}, due to the risk of \textit{transcription bugs}; that is, logic errors in the implementation of the specification.

Furthermore, it can be difficult to ensure in a large engineering organization that changes in code are accompanied by changes in the specification. \textit{eXtreme Modelling}, in which ``model and implementation are developed in parallel'' \cite{Gravell11ConcurrentDevelopmentOfModelAndImplementation}, has been proposed to prevent implementations and specifications from diverging.
Of the several guidelines included in this method, we attempted the following:
\begin{enumerate}[itemsep=-0.5ex]
  \item Multiple specifications model aspects of the system
  \item Specifications are written just prior to the implementation
  \item Specifications evolve with the implementation
  \item Tests are generated from the model and/or trace-checking verifies that test traces are legal in the specification
\end{enumerate}

Although eXtreme Modelling was not intended for existing code bases and specifications, we attempted to apply it to two of our existing systems. We intended to lay the groundwork to use this technique on future projects in which we would write the specification before the implementation, then keep the two in conformance as both evolved.

\subsection{Contributions}

We undertook two case studies simultaneously, using two separate code bases.
We attempted model-based trace-checking (MBTC) in the MongoDB Server.
One of our requirements was that the marginal cost of implementing MBTC for each future specification should be cheap, following an initial investment in trace-checking infrastructure.
After a great deal of effort, we determined that future specifications would not be easily trace-checked, and decided MBTC was impractical for the MongoDB Server.
We were impeded by the system's enormous code size, by its complex internal concurrency control, and by the oversimplifications we had made when writing our TLA\textsuperscript{+} specifications, before we knew they would be used for MBTC.
Concurrently, in MongoDB Realm Sync, a separate project with its own code base, we succeeded in using model-based test-case generation (MBTCG) to ensure conformance between an existing implementation, a newly written specification, and a subsequent implementation of the same specification in a different programming language.

This paper shows what factors made MBTCG successful and MBTC unsuccessful,
identifies lessons learned in both techniques, and suggests avenues of future research to make MBTC more practical.

We provide background to this research in Section \ref{sec:background}. 
Section \ref{sec:related_work} reviews related work. 
Section \ref{sec:model_based_trace_checking} describes our attempt at MBTC, and the challenges met. 
Section \ref{sec:model_based_test_case_generation} describes our success with MBTCG. 
Finally, Section \ref{sec:conclusions} describes our conclusions and avenues for future work.

\section{Background}
\label{sec:background}

\subsection{MongoDB Server}
\label{subsec:background_server}

The MongoDB Server is a distributed document database with features that are traditionally found both in NoSQL and Relational DBMSes.
It has support for flexible document schemas, as well as ACID compliant transactions and secondary indexes \cite{Kamsky19TPCCMongoDB}.
The MongoDB Server is typically deployed as a redundant group called a \textit{replica set}, which uses a protocol inspired by Raft \cite{Ongaro14Raft} to elect a leader node and replicate data changes to follower nodes.
Each node maintains a copy of the data and a durable log of operations, the \textit{oplog}.
Reads and writes offer multiple consistency and durability levels with increasingly strong guarantees \cite{Schultz19TunableConsistency, Tyulenev19CausalConsistencyMongoDB}.
The MongoDB Server has been under development for 12 years, with hundreds of thousands of lines of code, and dozens of active contributors at any time.

\subsection{MongoDB Realm Sync}
\label{subsec:background_realm}

MongoDB Realm is an offline-first database with a full-duplex synchronization protocol; a client is able to upload new changes to a centralized server without needing to download new changes from the server. 
(The MongoDB Server and MongoDB Realm are two distinct products with no shared code.)
MongoDB Realm's synchronization protocol, called MongoDB Realm Sync, uses \textit{operational transformation} (OT) to provide automatic conflict resolution and ensure all clients and the server converge to the same data \cite{Stigsen19RealmPatent}.

Each of these peers maintains a copy of the data and a durable log of operations, referred to as the \textit{state} and the \textit{history}, respectively. 
Since the client and server are not in constant communication, it is likely for a client to have already made subsequent changes to its data and history when it receives new changes from the server. The incoming changes from the server must be ``rebased'' on top of the client's current state and history. 
The client identifies the \textit{merge window} for the incoming changes as all client history entries since the last entry integrated by the server. 
These entries are causally unrelated to the incoming changes from the server. 
The OT algorithm transforms the incoming changes to produce a modified set of changes that are consistent for all peers. 
A \textit{last write wins} rule may be used to resolve the lack of a causal ordering for these operations. 
A similar procedure also occurs on the server when it receives changes from a client.


MongoDB Realm Sync has been under development for four years, with tens of thousands of lines of code, and three active contributors at any time.

\subsection{Testing and TLA\textsuperscript{+}}
\label{subsec:background_testing_tla}

To ensure that the MongoDB Server provides its advertised consistency guarantees with good performance, we run thousands of correctness and performance tests in our continuous integration (CI) system \cite{Daly19iChangePointMongoDB}.
Both the MongoDB Server and MongoDB Realm rely heavily on randomized testing.
For the MongoDB Server, tests randomly perturb the topology state and run randomized commands \cite{Guo17MongoDBFuzzTester}.

Engineers on MongoDB's Server-Replication Team wrote specifications to verify several concurrent and distributed protocols in the Server \cite{Schultz19BugsLife}.
We chose the formal modelling language TLA\textsuperscript{+}, introduced by Leslie Lamport \cite{Lamport99TLAPlus, Lamport02SpecifyingSystems}.
A TLA\textsuperscript{+} specification defines a set of variables expressing a system's state, and a set of actions which  modify this state.
The specification may include invariants that must hold at each state, as well as \textit{temporal logic} formulas to express properties of the system's execution over time.
TLA\textsuperscript{+} is a precise mathematical language; thus the properties of a TLA\textsuperscript{+} specification can be proven correct, or validated by a finite model checker which exhaustively explores its behaviors.
The most commonly used TLA\textsuperscript{+} model checker is TLC \cite{TLC}.

MongoDB chose TLA\textsuperscript{+} due to its success in industry and academic distributed systems research, open source implementation, and rich tooling.
Prior to this paper's research, TLA\textsuperscript{+} specifications for the MongoDB Server's replication protocol, its initial sync protocol, and some aspects of its hierarchical locking rules had already been written.
TLA\textsuperscript{+} has helped us design a complex, novel reconfiguration protocol, reproduce and fix multiple bugs, and verify liveness properties \cite{Schultz19BugsLife}.
Our success with TLA\textsuperscript{+} on the MongoDB Server's replication protocol led to its use for the MongoDB Realm Sync OT specification.

We plan to write specifications for future features and existing distributed protocols, to gain confidence in critical aspects of our system.
We use the same GitHub repository and follow the same code review and testing process with our models as we do for our tests and code.
We continuously model-check our TLA\textsuperscript{+} code alongside our correctness tests, to ensure that any changes to our specifications are correct, and to document how the model checker should be configured for each specification.

\subsection{MBTC and MBTCG}
\label{subsec:background_mbtc_mbt}

The eXtreme Modelling approach includes two testing methods that seemed promising for MongoDB: model-based trace-checking (MBTC) \cite{Jard83AnApproachToTestingSpecifications, MBTC} and model-based test-case generation (MBTCG) \cite{Utting06PracticalModelBasedTesting}. 

In MBTC, the behavior of the implementation is observed during testing or in production.
The implementation produces an execution trace that captures its sequence of state changes.
This trace is checked against the specification, either while the implementation is running or post-hoc, to verify that the trace is a permitted behavior of the specification.
The greater the variety of observed behaviors, the more confidence one has in the implementation's conformance.

In MBTCG, the specification is used to generate test inputs and expected outputs, either exhaustively or from a carefully selected subset \cite{Dick93AutomatingGenerationOfTests}.
A test framework provides the generated inputs to the implementation and checks whether it responds as specified.

\section{Related Work}
\label{sec:related_work}

Our work builds primarily on the eXtreme Modelling approach proposed by Gravell, et. al. \cite{Gravell11ConcurrentDevelopmentOfModelAndImplementation}, and closely related work by Augusto, et. al. \cite{Augusto03ValidatingBusinessSystems}. 
In these papers, the authors discuss the co-evolution of two e-business applications implemented in Java with several small models written in Promela and B.
They describe specification engineering methods; for example, holding regular meetings between modellers and implementers, and maintaining a dictionary to translate between names in the specification and the implementation.
Such techniques seem promising, but the authors analyze their effectiveness when practiced on an example application developed by a team of four academics, not on commercial software developed by dozens of engineers.
We have attempted to put the ideas of eXtreme Modelling into practice with complex commercial software, and using TLA\textsuperscript{+} and C++ instead of Java, B, and Promela.
We especially emphasize the use of MBTC and MBTCG to test that our implementations conform to our specifications.
To our knowledge, this is the most thorough attempt to implement eXtreme Modelling in industry.

Besides MBTC and MBTCG, a variety of methods have been proposed to ensure that an implementation conforms to its formal models.
Some projects use the same language for the specification and implementation \cite{KerGre99}, some mechanically generate implementation code from the specification \cite{Houhou17CodeGenerationFromSpecification}, others apply model-checking or formal verification directly to the implementation \cite{Chudnov18ContinuousFormalVerification, Holzmann04ModelDrivenVerification}, and yet others perform verified step-wise refinement from the high-level specification all the way down to the implementation \cite{Eiriksson95UsingFormalVerification}.
Any of these methods provides high confidence that the implementation conforms to the specification. 
However, they do not appear to be applicable to the MongoDB Server or to MongoDB Realm.
In the research we have reviewed, they are limited to small programs, either written in C or a specialized language such as Prolog; they have not been demonstrated with our implementation language, C++, nor on systems with code sizes as large as ours.
They cannot be used when writing new models for a system that has already been implemented.

Compared to the methods above, MBTC and MBTCG are more appropriate for eXtreme Modelling.
Considering MBTC first, however, we found that many of the specific uses of MBTC described in the literature are incompatible with eXtreme Modelling.

Ural et. al. in \cite{Ural84AutomatedTestingOfProtocolSpecifications} propose to begin with a high-level specification and refine it in stages, using MBTC to gain confidence that each stage's specification is equivalent to the previous stage.
This technique conflicts with two of the eXtreme Modelling guidelines: multiple high-level specifications cannot describe a single system, and co-evolution of specification and implementation is not practical with this technique, because one would have to repeat the effort of multi-stage refinement from top to bottom with each change.

Tasiran et. al. \cite{Tasiran03AlphaMicroprocessor} implement MBTC to check that a simulated microprocessor's behavior conforms to its TLA\textsuperscript{+} specification.
After each simulation step they invoke the TLC model checker to determine if the step is permitted by the specification.
But before invoking TLC there is a post-processing step implemented in 20,000 lines of C++.
This agrees with our experience: we also found that post-processing an execution trace is complex and requires significant engineering investment (Section \ref{subsec:mbtc_solution}).
The authors describe a sophisticated method to measure test coverage: they first map the specification's full state space to a smaller one by eliminating states that are ``qualitatively the same'' using two TLC features, \textit{symmetries} and \textit{views}. Once the specification state space is reduced they measure the portion of it covered by tests.

The eXtreme Modelling method advocates writing a specification and implementation simultaneously, but in our two case studies we began with a complete implementation. 
There is some discussion of modelling existing implementations in \cite{Newcombe2014UseOfFormalMethodsAmazon}.
Neubauer et. al. in \cite{Neubauer12AutomatedContinuousQualityAssurance} use machine learning to infer a formal model of an existing system from its observed behaviors.
They use MBTC to check if later execution traces of the system still conform to the inferred model; if not, either the system has a bug or the system behaved correctly and the model must be updated.
This work approaches the eXtreme Modelling development style, but an inferred model is quite different from one written by humans; the purpose of the former is solely to detect behavior changes in the implementation, the latter expresses the designers' intent.

The inverse of MBTC is MBTCG, in which a formal specification is used to generate test cases that assert the expected behavior of the implementation. In \cite{Dick93AutomatingGenerationOfTests, Legeard02AutomatedBoundaryTesting}, programs were written in Prolog to perform \textit{partition analysis} and automatically divide the input domain into equivalence classes, such that testing inputs within each class or near the boundary of multiple classes thoroughly explores the implementation's behavior. Dick and Faivre in \cite{Dick93AutomatingGenerationOfTests} automatically ordered sequences of operations to find the shortest test which covers all state transitions. In our MBTCG case study, instead of using partition analysis to reduce the number of generated test cases, it was feasible to exhaustively test \textit{all} combinations of operations subject to a constraint on the length of the initial array.

\section{Model-Based Trace-Checking}
\label{sec:model_based_trace_checking}

Because of our growing investment in TLA\textsuperscript{+} specifications of MongoDB Server protocols, we wanted to test whether our C++ implementation conforms to our specifications.
MBTCG did not seem appropriate for these protocols due to their non-determinism. 
A new leader can be elected at any time, for example, and followers can fall behind or catch up unpredictably. 
We could have injected pauses or other mechanisms into the implementation to force it to follow a sequence of steps that matches a generated test case.
This would require a large investment in test code, however, and it would so dramatically change the system's behavior under test it would cast doubt on the results.

Instead, we attempted to use MBTC.
Our proposed method would check execution traces obtained from our existing tests, and we would deploy our trace-checking system to a cluster of continuous integration servers.
This would permit us to rapidly develop both the specifications and the implementation, while receiving quick feedback about divergences between them.



\subsection{Implementing MBTC}
\label{subsec:mbtc_solution}

\begin{figure}[t]
{\centering
\includegraphics[trim=120pt 0 600pt 0,clip,width=0.7\linewidth]{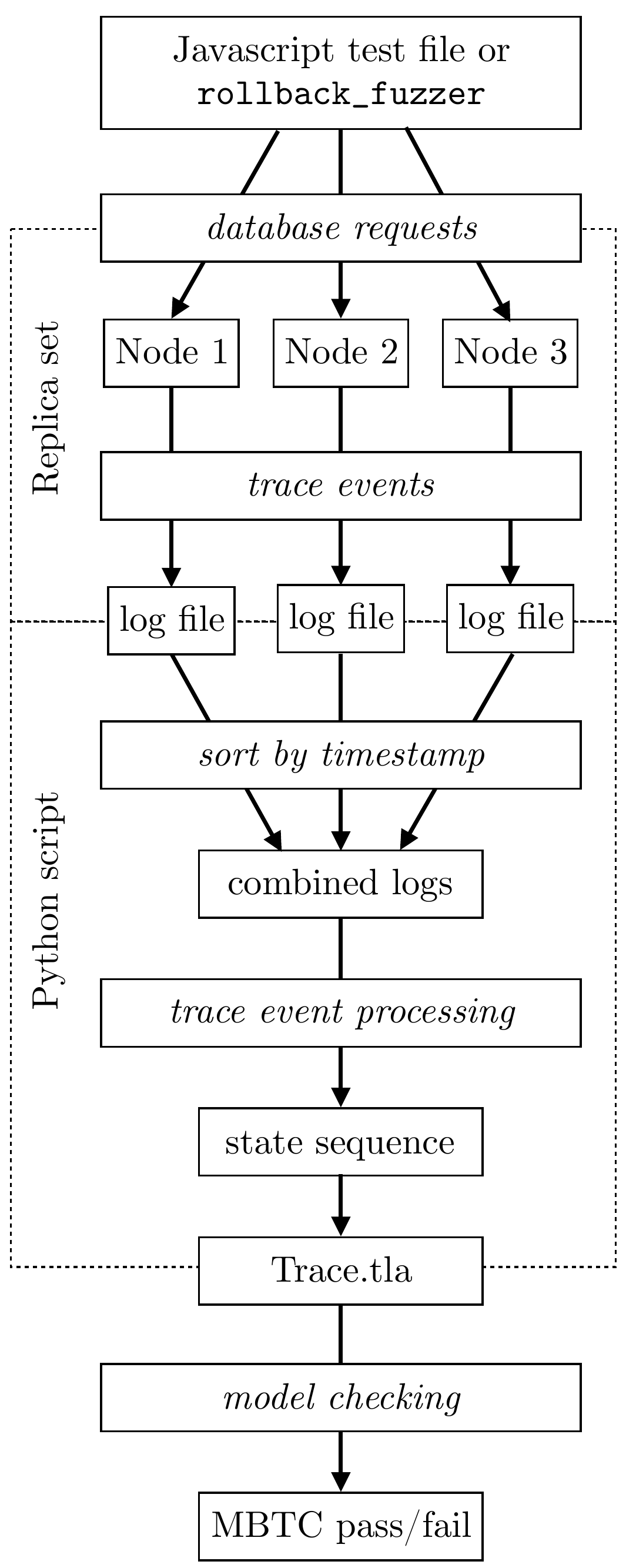}
\par}
\caption{MBTC data pipeline}
\label{figure:MBTC-pipeline}
\end{figure}

Prior to our research, there were 423 integration tests handwritten in JavaScript that target our replication protocol, and 23 randomized test suites that pause, disconnect, and terminate servers while the replica set is performing regular operations. 
Our final goal was to capture traces from all our tests and check them against all our specifications.

For our prototype we chose one specification, a 345-line TLA\textsuperscript{+} file called \texttt{RaftMongo.tla} in the MongoDB Server repository \cite{MongoGitHub}.
The specification's primary concern is to describe the gossip protocol by which nodes learn the \textit{commit point}: the newest oplog entry that has been replicated by a majority.
This specification is based on Diego Ongaro's TLA\textsuperscript{+} specification for Raft \cite{Ongaro14TLA+Raft}.
The same as in Raft, nodes in our specification can assume roles Leader or Follower. Time is divided into consecutively numbered election terms. A new term begins whenever a node calls for an election to choose a new leader.
One difference from Raft is that the MongoDB Server uses a pull protocol for replication, whereas Raft uses a push protocol; the MongoDB Server's follower nodes request oplog entries from the leader or other nodes, rather than the leader sending entries to followers.
\texttt{RaftMongo.tla} is very high-level with few state variables and invariants in order to focus on how the commit point is gossiped. 
Unlike Ongaro's work, our specification does not explicitly model messages passed between nodes.

Each node's state is modelled with four variables:

\begin{itemize}[itemsep=-0.5ex]
\item \texttt{role}: ``Leader'' or ``Follower''
\item \texttt{term}: Newest election term the node knows
\item \texttt{commitPoint}: Newest majority-committed oplog entry it knows
\item \texttt{oplog}: Contents of its oplog
\end{itemize}

\vspace{10 mm}

The specification has seven named state transitions:

\begin{itemize}[itemsep=-0.5ex]
\item \texttt{AppendOplog}: A node receives entries from any node
\item \texttt{RollbackOplog}: A node removes divergent entries
\item \texttt{BecomePrimaryByMagic}: A node is elected leader instantaneously---the election protocol is abstracted away
\item \texttt{Stepdown}: A leader becomes a follower
\item \texttt{ClientWrite}: A leader executes write operations
\item \texttt{AdvanceCommitPoint}: The leader advances the commit point
\item \texttt{UpdateTermThroughHeartbeat}: A node learns the election term from any node
\item \texttt{LearnCommitPointWithTermCheck}: A node learns the commit point from any node
\item \texttt{LearnCommitPointFromSyncSource\-Never\-Beyond\-Last\-Applied}: A node learns the commit point from a more up-to-date node
\end{itemize}



When we configure this specification with 3 nodes and constrain the model checker to at most 3 terms and oplogs up to 3 entries long, TLC successfully model-checks it and discovers 371,368 distinct states.
TLC validates an invariant that committed writes are not rolled back and a temporal property that the commit point is eventually propagated.

We wrote a C++ procedure \texttt{logTlaPlusTraceEvent}, enabled only in testing, which used our logging framework to emit as JSON the values of the four state variables above.
We found code locations in the MongoDB Server corresponding to the seven state transitions and added calls to this procedure.

MBTC with a distributed system requires a partial order of trace events; we achieved a strict order by running all processes on one machine and sleeping before logging the trace event.
Since all log messages include the current timestamp with millisecond precision, it was sufficient to sleep until the system clock's millisecond digit changed (Figure \ref{fig:millisecond_sleep}). (Jard et. al. describe a more general solution with vector clocks \cite{Jard94GeneralApproachToTraceChecking}, which was not necessary in our case.)

\begin{figure}[b]
\begin{verbatim}
PROCEDURE logTlaPlusTraceEvent(event) {
    /* Timestamp has millisecond precision */
    Timestamp beforeTime = getCurrentTimestamp()
    Timestamp afterTime = getCurrentTimestamp()
    while (afterTime == beforeTime) {
        sleep 1 millisecond
        afterTime = getCurrentTimestamp()
    }

    assert(afterTime > beforeTime,
           "Clock went backwards")

    log(event, afterTime)
}
\end{verbatim}
\caption{Pseudocode for \texttt{logTlaPlusTraceEvent}}
\label{fig:millisecond_sleep}
\end{figure}


We enabled tracing for our 423 handwritten JavaScript tests. 
Of these, 120 failed due to incompatibilities with tracing. 
(Section \ref{subsubsec:mbtc_impl_discrepencies} describes these incompatibilities.)
In total, these tests produced 42,262 trace events. 
We also selected one of our randomized tests, called \texttt{rollback\_fuzzer}, for MBTC. 
This test orchestrates network partitions which cause nodes to temporarily diverge, then to roll back writes and re-synchronize when the partitions are healed.
Random CRUD and DDL operations are run against leader nodes in the set to test that, with high probability, all combinations of operations and their behavior on rollback work consistently.
Nodes are also randomly restarted to test that clean and unclean restarts during rollback procedures do not cause data corruption.
A representative run of \texttt{rollback\_fuzzer} produced 2,683 trace events.

We wrote a Python script \cite{ReplTraceChecker} to post-process trace logs. 
The script merges the nodes' logs and sorts them by timestamp to obtain a sequence of trace events. 
Each event describes the state of only \textit{one} node at the moment after it executes a state transition.
In order to construct a sequence of states describing the \textit{entire} replica set, the script begins with a known initial state, combines it with the first trace event to determine the next state, and so on. 
The logic to combine a current state $S$ with a trace event $E$ from node $N$ to produce next state $S'$ is as follows:

\begin{itemize}[itemsep=-0.5ex]
\item \texttt{role}: The script assumes there are never two leaders at once (although this is unrealistic, see Section \ref{subsubsec:mbtc_impl_discrepencies}). Thus if the role of $N$ in $E$ is ``Leader'' then its role in $S'$ is ``Leader'' and all others' roles in $S'$ are set to ``Follower''. If $N$ had role ``Leader'' in $S$ and now has role ``Follower'' in $E$, its role in $S'$ is ``Follower'' and the other nodes' role values are unchanged. 
\item \texttt{term}: The term of $N$ in $S'$ is set to its term in $E$. The other nodes' terms are unchanged.
\item \texttt{commitPoint} and \texttt{oplog}: Similar to term.
\end{itemize}

\begin{figure}[b!]
\includegraphics[height=32em]{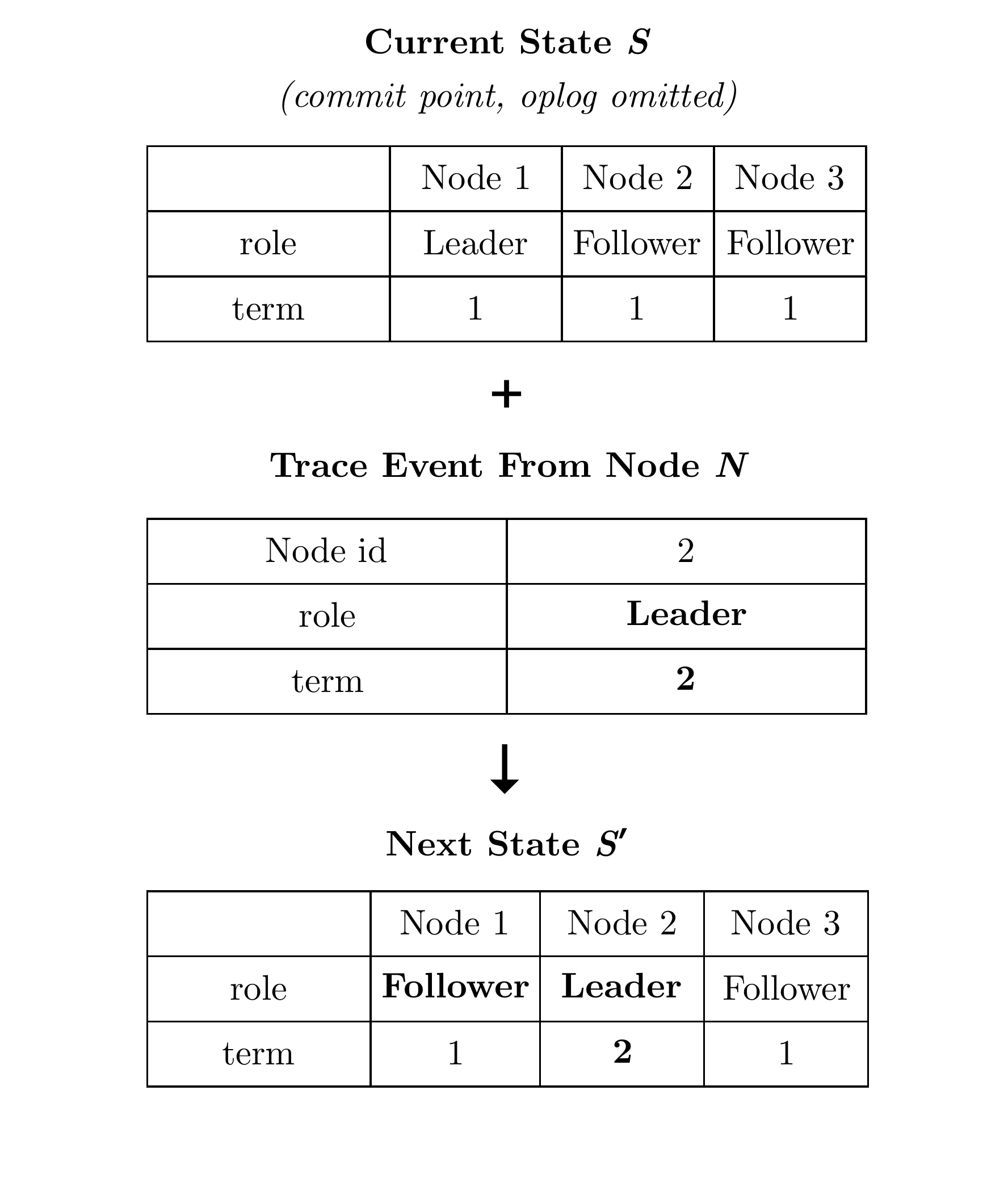}
\caption{Trace Event Processing}
\label{figure:event-processing}
\end{figure}

For example, in Figure \ref{figure:event-processing}, the replica set's current state has Node 1 as the leader in term 1. 
The script processes a trace event from Node 2 announcing it has become leader in term 2, arriving at the next state.

\begin{figure}
\includegraphics{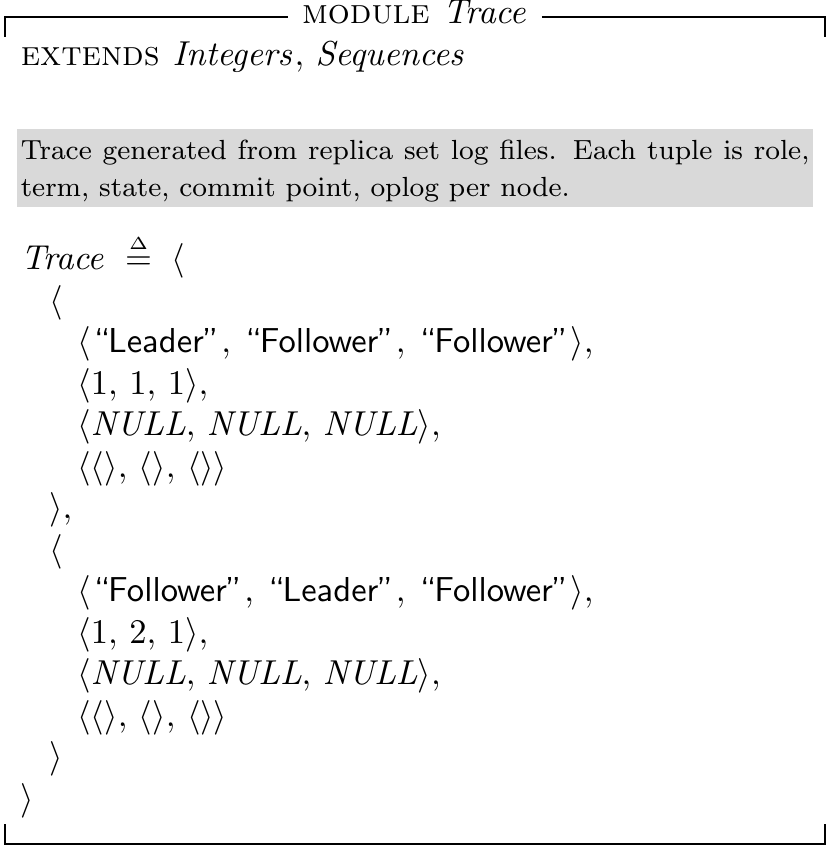}
\caption{State sequence as TLA\textsuperscript{+} tuple (simplified)}
\label{fig:state-sequence}
\end{figure}

Once the script constructs this sequence of states, it implements MBTC following a method proposed by Pressler \cite{Pressler18VerifyingSoftwareTracesTLAPlus}: it generates a TLA\textsuperscript{+} module called \texttt{Trace.tla} which includes the sequence of states (Figure \ref{fig:state-sequence}), and uses TLC to check that the sequence is permitted by the \texttt{RaftMongo.tla} specification.
The entire data pipeline can be seen in Figure \ref{figure:MBTC-pipeline}. 
It was implemented by two engineers in 10 weeks.

\begin{center}
\begin{tabular}{ | m{11em} | m{5em}| m{6em} | } 
\hline
Task & Effort & Lines of Code \\  
\hline
Event tracing & 4 weeks & 570 C++ \\ 
Update RaftMongo.tla & 3 weeks & 252 TLA\textsuperscript{+}\\ 
Python post-processor & 3 weeks & 484 Python \\ 
\textbf{Total} & \textbf{10 weeks} & \\
\hline
\end{tabular}
\end{center}

\subsection{Analysis}
\label{subsec:mbtc_analysis}


We had intended to trace-check all of our specifications against traces from all of our handwritten and randomized tests, deploy the trace-checker to our continuous integration system, and measure accumulated state space coverage over all tests. 
We had hoped that the marginal cost of checking each specification after the first would decrease.
Had we achieved this, we would have built much of the test infrastructure required for eXtreme Modelling.
However, we applied trace-checking to only 5 handwritten tests and one randomized test. 
Only one handwritten test generated traces that passed the trace-checker; the other 4 produced traces that violated the specification due to two implementation discrepancies (see the discussion of initial sync and term in Section \ref{subsubsec:mbtc_impl_discrepencies}). 
We did not deploy to continuous integration nor measure coverage. 
The effort to implement MBTC proved so costly, we abandoned the project after 10 weeks of engineering effort. 
We faced complexities with concurrency and locking, discrepancies between our models and implementation, and incomplete support in TLC. 
With further effort we might have achieved MBTC for one specification, but we estimated that the marginal cost of checking each additional specification would approach the cost of the first.

In the following sections we describe our issues trace-checking \texttt{RaftMongo.tla}, argue that additional specifications would be just as costly, and offer advice for future implementers of MBTC.

\subsubsection{Visibility, hierarchical locking}
\label{subsubsec:mbtc_locking}

\textbf{Visibility}: When we began to add tracing to the MongoDB Server, we realized each trace event must be logged after it has occurred, but \textit{before} the change is visible to other nodes. 
For example, when a leader receives a write from a client application and creates an oplog entry, it must log the \texttt{ClientWrite} event after the entry appears in its own oplog, but before any followers can replicate the entry and log an \texttt{AppendOplog} event. 
If a follower logged an \texttt{AppendOplog} event with an earlier timestamp than the leader's \texttt{ClientWrite} event, the trace would violate the causal relationship described in the specification. 
To ensure each state change was logged before it became visible to other nodes, our logging code had to hold several locks.

\textbf{Hierarchical locking}: Formal specifications of distributed systems algorithms, such as Raft, model a concurrent system of interacting processes, but they typically model each process as single-threaded. 
Production database systems such as the MongoDB Server, however, almost always have high intra-process concurrency and employ some degree of hierarchical locking \cite{Gray76SharedLocks}. 
The MongoDB Server specifically combines hierarchical locking with storage-engine Multi-Version Concurrency Control (MVCC), C++ latches, and higher level concurrency control primitives like futures.
It may not be feasible to log a consistent snapshot of such a process's state at the moment of a trace event.

One approach we considered, was to only log the state that actually changed with each event, and use post-processing to fill in the missing state.
We already used this approach to only log one process's state at a time rather than attempting to snapshot multiple processes.
We chose not to do this for fear of masking bugs if we incorrectly post-processed trace events.
In hindsight it may have made trace logging more cost effective with little loss in value.


To implement MBTC for \texttt{RaftMongo.tla}, a node must include the contents of its oplog with each trace event, but acquiring the locks to obtain a snapshot of the oplog proved difficult. 
Suppose our \texttt{logTlaPlusTraceEvent} procedure (Figure \ref{fig:logTlaPlusTraceEvent}) must acquire locks A, B, and C, in that order, to read the oplog.
Consider a procedure \texttt{becomeLeader} that acquires locks A and C, then changes the node's role to Leader. 
If we add a call from \texttt{becomeLeader} to \texttt{logTlaPlusTraceEvent}, the latter must acquire lock B, but this is the wrong order and risks deadlocking with other threads.
If \texttt{becomeLeader} first \textit{dropped} lock C, then acquired locks B and C, that would be the correct order.
But dropping lock C could allow a thread servicing an external request to communicate the node's new role to another process, thus making the role change \textit{visible} and violating the visibility rule described above.
General solutions are unpalatable: callers of \texttt{logTlaPlusTraceEvent} could be responsible for acquiring all the locks it will need, but this encodes intimate knowledge about \texttt{logTlaPlusTraceEvent} into its callers, and it significantly alters the system's behavior under test.

\begin{figure}
\begin{verbatim}
PROCEDURE becomeLeader() {
    acquire Lock A
    acquire Lock C
    role := Leader
    logTraceEvent()
}

PROCEDURE logTlaPlusTraceEvent() {
    /* Wrong acquisition order if called by
       becomeLeader, risks deadlock */
    acquire Lock A if not yet acquired
    acquire Lock B if not yet acquired
    acquire Lock C if not yet acquired
    read oplog
    /* Remainder of procedure as in Fig. 2 */
}
\end{verbatim}
\caption{Pseudocode for a node becoming a Leader}
\label{fig:logTlaPlusTraceEvent}
\end{figure}

Solving both the visibility and the locking challenges, for each of the seven named state transitions in \texttt{RaftMongo.tla}, was the single most difficult aspect of our MBTC implementation, costing roughly a month of engineering effort. 
Eventually we discovered code locations that obeyed each transition's visibility requirements.
We managed to avoid the need for acquiring locks out of order by exploiting the MongoDB Server's MVCC features: our storage engine can serve reads from a stale snapshot of the oplog, instead of locking the oplog to read its current contents.
Fortunately, in each location where \texttt{logTlaPlusTraceEvent} could not acquire the locks needed to read the current oplog, a stale snapshot was permitted by the specification.
In each code location where the specification required the most recent oplog, on the other hand, it was possible to lock it.


In contrast to our difficulties adding tracing to the MongoDB Server, Gravell et. al. \cite{Gravell11ConcurrentDevelopmentOfModelAndImplementation} emphasize how easily they added tracing.
We believe their task was easier because 1) their application was a small prototype, 2) they had deliberately written their specification and implementation to closely correspond (see Section \ref{subsubsec:modelling_for_trace_checking}), and 3) their application did not support the complex intra-process concurrency of a typical database.
We were surprised at the difficulty of adding tracing to the MongoDB Server, both because it had been easy for Gravell et. al. and because we had not anticipated how visibility and locking would complicate our code.
We expect any MBTC implementation for a concurrent database would encounter similar complexities with hierarchical locking and visibility of state changes.

\subsubsection{Implementation discrepancies}
\label{subsubsec:mbtc_impl_discrepencies}

As soon as we checked our tests' execution traces against \texttt{RaftMongo.tla}, we encountered violations.

\textbf{Arbiters}: MongoDB Server nodes can be configured as \textit{arbiters} which vote in elections but have no data. 
\texttt{RaftMongo.tla} does not model arbiters and we did not implement tracing for them; arbiters crash when tracing is enabled.
It was trivial to skip tests that use arbiters, but this gave us a hint that our specification was an oversimplification of the real system, and would not be suitable for MBTC.

\textbf{Initial sync}: The second discrepancy was in the behavior of \textit{initial sync}, the process a newly-added replica set member uses to obtain a copy of the data and oplog.
Trace-checking the output of \texttt{rollback\_fuzzer} immediately reproduced a known violation of the specification: our implementation considers an oplog entry to be majority-committed once a quorum of nodes, including initial-syncing nodes, has replicated it.
However, initial-syncing nodes should not be considered quorum members because their oplog entries are not durable until initial sync completes.
When we first wrote \texttt{RaftMongo.tla} we modelled the correct behavior.
We did not have MBTC in mind, so we deliberately wrote an idealized specification with the intention to eventually bring our implementation into conformance.
This violation increased our confidence in trace-checking.
It illustrated how well MBTC works in practice, and had we not already known about this behavior, it would have been a significant discovery.
However, it made our MBTC project nearly worthless: the violation came only 4 steps from the trace's start and left the remaining 2,683 steps unchecked.

To permit deeper trace-checking, we could 1) fix the implementation of initial sync sooner than planned, 2) avoid triggering the non-conforming behavior in testing, 3) update the specification to match the flawed implementation, or 4) post-process the traces to simulate a conformant implementation.
Fixing the implementation would require substantial work and is already scheduled for a future release.
Instead, we chose solution 2 and modified \texttt{rollback\_fuzzer} so all followers were fully synced before the test began any writes.

\textbf{Two leaders}: As in Raft, MongoDB's election protocol permits two leaders to co-exist briefly.
However, we chose in \texttt{RaftMongo.tla} to assume there is at most one leader, since the election protocol is not the specification's main concern.
To avoid trace-checking failures, we again chose solution 2, and avoided tests exhibiting two leaders.

\textbf{Term}: Other discrepancies required solution 3, updating the specification. 
Since the election protocol was not its focus, \texttt{RaftMongo.tla} originally modelled the election term as a single global number known by all nodes.
In reality, election terms are gossiped among nodes and each learns the new term at a different time.
MBTC required us to make the specification more complex to match reality. 
We chose to update the specification to resolve many other discrepancies, as well: over the course of our research we added or changed 252 of the 345 lines of TLA\textsuperscript{+} in \texttt{RaftMongo.tla}, costing three weeks of effort.

\textbf{Copying the oplog}: In several cases we chose solution 4, post-processing the logs to simulate conformance. For example, in \texttt{RaftMongo.tla}, when a node performs an initial sync, it copies the leader's entire oplog.
In the implementation, a new node copies only recent entries.
The \texttt{RaftMongo.tla} behavior more closely matches the Raft protocol that inspired it, and makes it simpler to express invariants such as ``an entry is never rolled back after it is replicated to a majority of nodes' oplogs.''
We resolved the discrepancy by adding logic to our Python script that filled in the missing entries while it generated the state sequence.
Such an intrusive modification made us concerned that a mistake in the Python script might mask a harmful transcription bug in the MongoDB Server code.
On the other hand, choosing solution 3 and adding detail to the specification might explode the state space and make model-checking intractable.
At the least, it would make the specification harder to read.

Each discrepancy we discovered through MBTC required us to judge which of the four solutions to employ. 
Ideally, when an organization practices eXtreme Modelling, they resolve each MBTC failure either by fixing the implementation or, if the failure is not a bug, by updating the specification.
In practice, we often concluded it was expedient to work around violations.
We chose to avoid the non-conforming behavior (solution 2) or to simulate a conformant behavior (solution 4) depending on which took less effort and least undermined our confidence in the test.
Unfortunately, our choices were based on estimates and speculation.

We might have resolved some discrepancies with a refinement mapping \cite{Abadi88ExistenceOfRefinementMapping} from \texttt{RaftMongo.tla} to a lower-level specification which would be used for MBTC. In the "copying the oplog" discrepancy above, it seems possible to use a history variable \cite{Lamport2018AuxiliaryVariables} to bridge the gap between our specification and a more realistic one. In other cases (two leaders, term gossiping) either no refinement mapping exists or it would require great effort to find.

\subsubsection{Modelling for trace-checking}
\label{subsubsec:modelling_for_trace_checking}

There appears to be a conflict between writing a specification for documenting and model-checking a design, and writing one that can be trace-checked.

We tried to implement MBTC with models that had been written for documentation and model-checking, but we found that MBTC is only practical if the specification is written with MBTC in mind.
If we began again, we would rewrite our specifications to closely correspond to the implementation.
The specification's major components would be structured similarly to the implementation's, and critical sections of the implementation would correspond with single actions in the specification.
In the case of multi-step events such as elections, we would model them with multiple actions, instead of expressing them as instantaneous single actions as we did with \texttt{BecomePrimaryByMagic} in \texttt{RaftMongo.tla}.
We would faithfully model the flaws in the implementation if we did not plan to fix them immediately.

We would rewrite our specification to model events that are easily observed in the implementation: for example, we might model protocol messages, akin to the original Raft specification.
We would try to avoid modelling state that is difficult to snapshot, especially if it is protected with complex locking.
If some varibles in the specification cannot be logged by the implementation, Pressler proposes a refinement mapping technique \cite{Pressler18VerifyingSoftwareTracesTLAPlus} in which TLC checks whether there is any sequence of values for the missing variables that would permit the trace to match the specification.

Most of these changes in our modelling style would be unobjectionable.
However, there is the danger that a less abstract specification has a state space too large to model-check in reasonable time.
The limited changes we made to \texttt{RaftMongo.tla} increased the state space from 42,034 states to 371,368, and model-checking time from 2 seconds to 14 minutes.
To complete our MBTC project would have required a more detailed specification, or applying Pressler's refinement mapping technique, both of which would threaten to explode the state space even more.

\vspace{10 mm}

\subsubsection{Tooling and TLC}
\label{subsubsec:mbtc_tla_tlc}

Tool support for MBTC with TLA\textsuperscript{+} and TLC is a work in progress. 
Pressler's method worked well to check traces of hundreds of events, but for thousands of events it was impractically slow. 
Pressler proposed, and Markus Alexander Kuppe has begun to implement, features to check long traces by bypassing the TLA\textsuperscript{+} parser in favor of a special-purpose Java extension to TLC \cite{TLAPlusIssue413}. 
MBTC with TLA\textsuperscript{+} will be more convenient once these features are released, with publicly available example code. 
Another missing feature is the ability to combine state-space coverage reports over multiple TLC executions on different traces, which would permit engineers to calculate the total coverage achieved by deploying MBTC to continuous integration.

\subsubsection{Marginal cost of trace-checking a specification}
\label{subsubsec:marginal_cost}

Our experience with \texttt{RaftMongo.tla} convinced us that trace-checking additional specifications would be nearly as costly as the first.
For example, we imagined a scenario in which we had succeeded at trace-checking \texttt{RaftMongo.tla} and we moved on to \texttt{Locking.tla} \cite{LockingTLA}, which models aspects of the MongoDB Server's lock hierarchy.
Once again we would begin with a highly abstract specification written for documentation and model-checking, and we would repeat the effort of adding detail and resolving discrepancies to make the specification suitable for MBTC.
The state variables of \texttt{Locking.tla} are disjoint from those modelled by \texttt{RaftMongo.tla}, so there would be little code reuse in the tracing implementation or the post-processing script.
The locking specification applies only to one process, not to an entire replica set, so the log post-processing script would be simpler for \texttt{Locking.tla} than for \texttt{RaftMongo.tla}; however, this difference would further preclude code reuse.

We canceled the MBTC project because we had achieved disappointingly little utility from our efforts, and our revised estimate of the marginal cost made it clear that the project was not worthwhile.




\section{Model-Based Test-Case Generation}
\label{sec:model_based_test_case_generation}

Our second case study involved a distinct product, MongoDB Realm Sync.
Architectural decisions have led us to re-implement the MongoDB Realm Sync server in Golang while keeping the clients in C++. Originally, a shared C++ code base ensured the client's and server's behavior were identical. Now that there are two implementations, it is paramount they resolve conflicts identically. Designing operational transformation (OT) functions that ensure peers converge to the same data is already challenging and error-prone \cite{Imine03ProvingCorrectnessOfTransformationFunctions, Imine06FormalDesignAndVerificationOfOT, Randolph13OnConsistencyofOTApproach}; implementing it the same way twice is even more difficult.

Model-based test-case generation (MBTCG) \cite{Gravell11ConcurrentDevelopmentOfModelAndImplementation} provided means to confidently assess whether parity had been achieved while both the C++ and Golang code bases were under active development. We wrote a new TLA\textsuperscript{+} specification for this purpose and used the state space explored by the TLC model checker to generate C++ and Golang test cases for aspects of the OT algorithm. These test cases can be automatically generated and run after any change to the model, and ensure that both the C++ and Golang implementations make appropriate changes to maintain equivalence.


MongoDB Realm Sync has 19 distinct operations which can be performed on a group of tables, an individual table, an object, or a list of values. These include operations such as deleting a table, creating a new object in a table, setting the property of an object, and inserting a new element into a list. Each type of operation must define how it merges with all other types of operations. This yields $19 (19 + 1) / 2 = 190$ merge rules that must be defined, with the remaining $19^2 - 190 = 171$ merge rules inferred by symmetry. Approximately three-quarters of the merge rules have trivial implementations where the incoming operation is applied unchanged by both peers.

The most complex merge rules are for the six array-based operations:

\begin{itemize}[itemsep=-0.5ex]
  \item \texttt{ArraySet}: replacing the value of an existing element
  \item \texttt{ArrayInsert}: inserting a new element at a position within the list, or growing the size of the list by one
  \item \texttt{ArrayMove}: moving an element from one position to another
  \item \texttt{ArraySwap}: swapping the position of two elements
  \item \texttt{ArrayErase}: removing an element from the list
  \item \texttt{ArrayClear}: removing all of the elements from the list
\end{itemize}

The $6 (6 + 1) / 2 = 21$ merge rules for these six operations are implemented in approximately 1,000 lines of C++.

\subsection{Writing the TLA\textsuperscript{+} specification}
\label{subsec:mbtcg_tlaplus}

MongoDB's Realm Sync Team wrote a TLA\textsuperscript{+} specification for these array-based operations to assess the soundness of the existing C++ implementation and to exhaustively generate test cases for a new Golang implementation.



\begin{figure}
\includegraphics{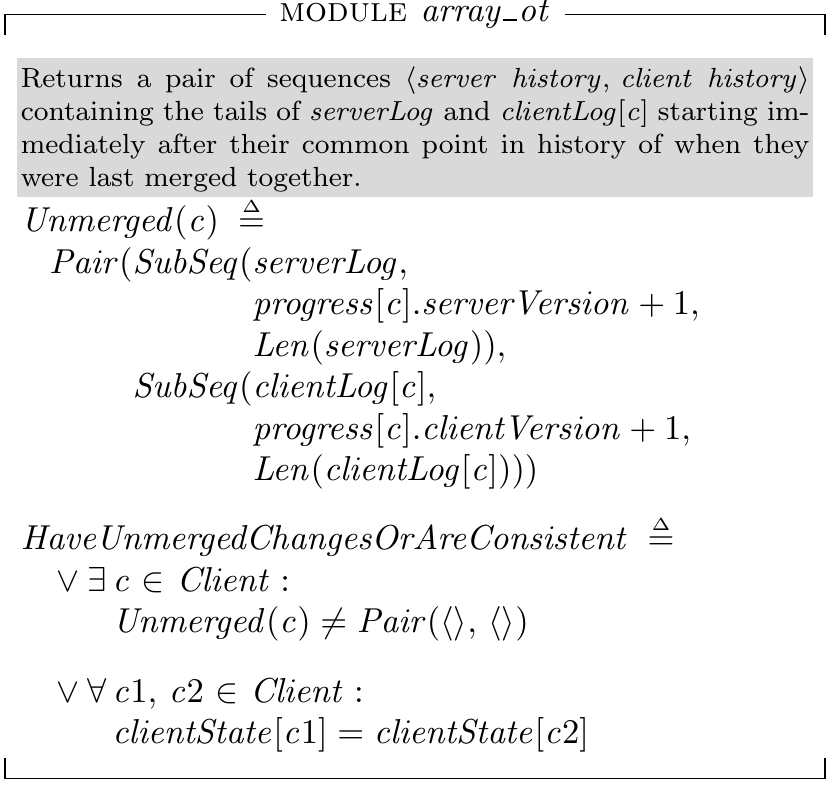}
\caption{TLA\textsuperscript{+} invariant for MongoDB Realm Sync}
\label{fig:tlaplus_realm_sync_invariant}
\end{figure}

Writing the TLA\textsuperscript{+} specification took approximately 40 hours over the course of two weeks. It was written by copy-pasting the C++ code and manually updating the syntax to be valid TLA\textsuperscript{+}. The merge rule for two \texttt{ArrayInsert} operations was written first along with the \texttt{HaveUnmergedChangesOrAreConsistent} invariant (see Figure \ref{fig:tlaplus_realm_sync_invariant}).
The merge rules for each subsequent operation type were then added and checked with TLC after each addition.
TLC found transcription errors multiple times during this process. The counterexample produced was manually analyzed by associating the branches taken in the series of if-statements in the C++ code with the nested \texttt{IF} and \texttt{CASE} expressions taken in the TLA\textsuperscript{+} specification.

This transcription process was challenging in two ways. First, the C++ code relies on mutable variables for expressing how the incoming operations must be modified to be correctly applied by the non-originating peer, whereas TLA\textsuperscript{+} variables cannot be mutated in the same way. Second, the order in which clients perform operations locally and then later merge with the server doesn't affect their final state, but, if left unconstrained, would quickly lead to a state space explosion.

\begin{figure}
\includegraphics{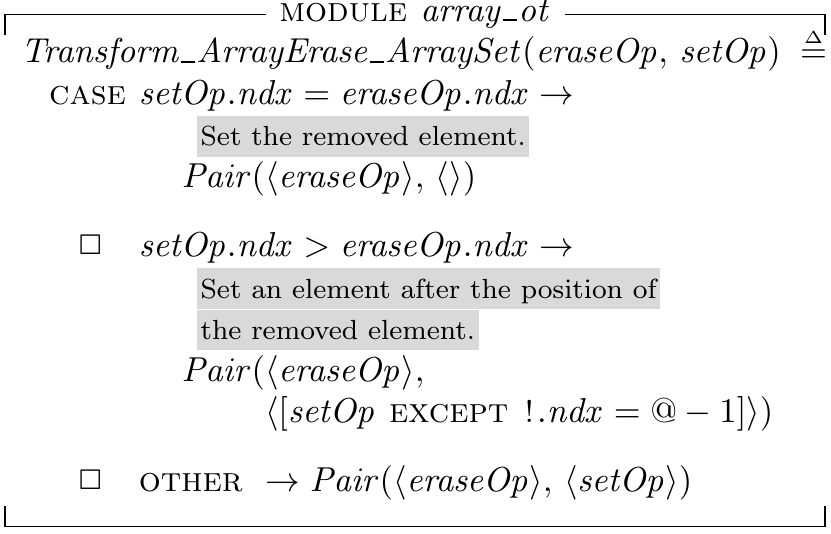}
\caption{TLA\textsuperscript{+} code for merging \texttt{ArrayErase} and \texttt{ArraySet} operations}
\label{fig:tlaplus_erase_set_merge}
\end{figure}

\subsubsection{Handling C++ mutability and flow control}

TLA\textsuperscript{+} is not a general-purpose programming language. \emph{Operators} in TLA\textsuperscript{+} (which behave most closely to a function in a general-purpose programming language) cannot reassign an argument or have multiple statements as in C++ functions.
While PlusCal offers an imperative syntax, it only supports reassigning state variables---and not arguments within an operator---because it transpiles to TLA\textsuperscript{+}.
As shown in Figure \ref{fig:cpp_erase_set_merge}, the C++ variable \texttt{set\_op} is mutated to produce the version of these operations to be applied on the non-originating peer. This mismatch between the two languages makes it more difficult to formalize a C++ source of truth into a TLA\textsuperscript{+} specification, especially in the presence of nested if-statements. Additionally, a series of multiple if-statements must be transcribed to explicitly define all branch combinations that may or may not be taken consecutively. Fortunately, the TLC model checker was readily able to catch human transcription errors as safety violations. Some examples of transcription errors were forgetting to substitute the updated index number in later comparisons, or entirely forgetting to embed a deeply-nested or later occurring if-statement as another nested \texttt{IF} or \texttt{CASE} expression in TLA\textsuperscript{+}.

\begin{figure}
\begin{verbatim}
DEFINE_MERGE(ArrayErase, ArraySet)
{
  if (same_array(erase_op, set_op)) {
    // LCOV_EXCL_STOP
    if (set_op.ndx == erase_op.ndx) {
      // CONFLICT: Update of a removed element.
      // RESOLUTION: Discard the ArraySet
      // operation.
      set_op.discard();
    }
    else if (set_op.ndx > erase_op.ndx) {
      set_op.ndx -= 1;
    }
    // LCOV_EXCL_START
  }
}
\end{verbatim}
\caption{C++ code for merging \texttt{ArrayErase} and \texttt{ArraySet} operations}
\label{fig:cpp_erase_set_merge}
\end{figure}

\vspace{10 mm}

\subsubsection{Constraining the state space}

The state constraint was progressively refined as the TLA\textsuperscript{+} specification was being implemented in order to reduce the run-time required by the TLC model checker. One simplification was a \texttt{MergeAction} operator which simultaneously uploads any changes from the client and downloads any changes from the server. Modelling these as distinct actions would only have been beneficial when explicitly modelling time. Instead, the model's purpose is to generate C++ and Golang test cases where clients perform a sequence of operations starting from the same initial array. A client performing a sequence of operations that are causally dependent on another client's sequence of operations is equivalent (in terms of both its state and history) to the former client performing the concatenation of the two sequences on the same initial array. However, a longer sequence of operations necessarily has more combinations and would lead to a state space explosion. We therefore shifted to focus on varying the initial array and restricted all clients to performing a single operation. Since the TLC model checker would explore all combinations of the clients performing a single operation, starting with a sufficiently large initial array would still enable the TLC model checker to exercise every case in the merge rules.

We artificially constrained the state space to have clients perform and merge operations in ascending order of their ID to avoid exploring redundant states. In particular, it doesn't matter which order clients 1 and 2 perform operations locally until they communicate with the server. We aren't able to define the clients as a symmetry set because the ID is used to order operations when their timestamps are equal. As mentioned previously, the specification doesn't model time, so we need at least three clients to capture a client merging both with an earlier operation and with a later operation.
Using the minimum of three clients ensures we don't increase the state space unnecessarily.
The TLC model checker was run with constraints of three clients each performing a single operation on an array already containing three elements.

\subsubsection{Results from the TLC model checker}

In addition to detecting safety violations caused by transcription errors, the TLC model checker also encountered a \texttt{StackOverflowError} due to a case in the merge rule for the \texttt{ArraySwap} and \texttt{ArrayMove} operations leading the merge function to never terminate. This issue was found to also exist in the C++ code; the bug had faithfully been transcribed from C++ to TLA\textsuperscript{+}. It was surprising to discover a bug in the C++ implementation due to the maturity of the product and the amount of randomized testing it had undergone.
The discovery of this issue became the deciding factor to not support a dedicated \texttt{ArraySwap} operation in the new Golang server implementation and to deprecate it in the existing C++ implementation. The \texttt{ArraySwap} operation was therefore excluded from testing for the remainder of this experiment.

Even if the model checker does not find a violation of the \texttt{HaveUnmergedChangesOrAreConsistent} invariant, this does not imply that the specification accurately describes the C++ implementation.
For example, a trivial way to achieve convergence (at the cost of user-intent preservation) would be to have every merge rule clear the contents of the array. We must test that the behavior of the specification conforms to the C++ code.

\subsection{Generating C++ test cases}
\label{subsec:mbtcg_cpp}

The state space explored by the TLC model checker provides an exhaustive set of test cases for how a given pair of operations should be transformed when applied by the non-originating peer. The TLC model checker supports writing the graph of all reachable states to a GraphViz DOT file, where each edge is labeled with the action taken to progress from one state to the next. We wrote a Golang program to parse this file and generate C++ test cases composed of (1) the initial array, (2) the operations each client performed, (3) the transformed version of the operations each client applied after merging the original operations from the other clients, and (4) the final state of the array. The Golang program then produces a C++ program which uses MongoDB Realm Sync's unit test framework. The \texttt{fixture.sync\_all\_clients()} call takes the place of performing the merge action for all the clients (in the same order the TLA\textsuperscript{+} specification performs the merge).

\begin{figure}
\begin{verbatim}
TEST(Transform_Node__6971023528664242108)
{
  size_t num_clients = 2;
  TransformArrayFixture fixture{
    test_context, num_clients, {1, 2, 3}};

  fixture.transaction(0, [](TableRef array) {
    array->set_int(0, 2, 4);
  });
  fixture.transaction(1, [](TableRef array) {
    array->remove(1);
  });

  fixture.sync_all_clients();
  fixture.check_array({1, 4});

  fixture.check_ops(0, {ArrayErase{1}});
  fixture.check_ops(1, {ArraySet{1, 4}});
}
\end{verbatim}
\caption{C++ test case for merging \texttt{ArrayErase} and \texttt{ArraySet} operations}
\label{fig:cpp_test_erase_set_merge}
\end{figure}

For an initial array containing three elements and with three clients each performing a single operation, the Golang program generated 4,913 C++ test cases.
The coverage of the specific sections of the C++ code was measured by adding in \texttt{LCOV\_EXCL\_STOP} and \texttt{LCOV\_EXCL\_START} comments within the block of the \texttt{same\_array()} condition. This was done because the TLA\textsuperscript{+} specification was designed to only exercise cases where the clients act on the same array.
The 36 handwritten C++ test cases covered 18 of the 86 branches (21\%) within the merge rules for the array-based operations.
This low percentage is excusable because we also rely on fuzz testing.
The \texttt{fuzz-transform} test executable, which uses the AFL fuzzer \cite{AFLHomepage} to produce randomized inputs that are then mapped to randomized operations, covered 79 of 86 branches (92\%) after approximately 8 million total executions.
The generated C++ test cases covered all 86 branches (100\%).


Achieving 100\% branch coverage, with all the generated C++ test cases passing, gives us confidence that the TLA\textsuperscript{+} specification conforms to the C++ code.
There may still be programming bugs in the C++ implementation stemming from undefined behavior.
However, for the well-formed inputs from the TLC model checker, we can be certain the OT algorithm converges and the C++ and Golang implementations of the merge rules for the array-based operations always agree.
It was worthwhile for a single engineer to spend one month writing the TLA\textsuperscript{+} specification and implementing the C++ test case generator.


\begin{center}
\begin{tabular}{ | m{11em} | m{5em}| m{6em} | } 
\hline
Task & Effort & Lines of Code \\  
\hline
array\_ot.tla & 2 weeks & 795 TLA\textsuperscript{+} \\
C++ test case generator & 2 weeks & 755 Golang \\
\textbf{Total} & \textbf{4 weeks} & \\
\hline
\end{tabular}
\end{center}

\section{Conclusions}
\label{sec:conclusions}

MBTC was impractical for our system because we lacked a close correspondence between the MongoDB Server implementation and \texttt{RaftMongo.tla}.
This was partly by necessity: the MongoDB Server is hundreds of thousands of lines of code; the replication protocol alone is tens of thousands of lines.
Any close correspondence with the MongoDB Server's implementation would have made the specification overly complicated for its purpose, and likely it would have inflated the state space beyond what is feasible to model-check.
In the absence of a close correspondence, MBTC requires significant post-processing of traces. 
We learned that in a system of the MongoDB Server's complexity, post-processing should be used early and often.
Our trace-logging code was hampered by the Server's intra-process concurrency control.
Future engineers in our situation should consider logging only those state variables the implementation can access at the moment it executes a state transition.
A post-processing script can fill in the missing variables, so long as there is some reliable way to infer their values post-hoc.
For example, it might be guaranteed that the missing variables' values have not changed since the last time they were logged.
Bugs in post-processing could mask implementation bugs, so it is critical to think about what kinds of bugs the MBTC effort aims to catch and what kinds are targeted by other tests.

An avenue we did not explore is to generate traces from implementation modules running in a unit test framework, rather than an integration test of the entire multi-process system, as we did.
Modules are simpler than the whole, and non-determinism can be more easily eliminated, for example by simulating the system clock. 
The difficulties we experienced with visibility and hierarchical locking would likely be more manageable.
By testing modules in isolation, one could sacrifice realism in exchange for implementing MBTC cost-effectively.
Especially if there are multiple specifications modelling aspects of the system, as in our case, it would be appropriate to pair a specification with each module.

Regardless, MBTC in a system of the Server's scale will require significant effort for every specification.
No initial infrastructure investment can make the marginal cost of trace-checking additional specifications cheap.

\balance

MBTCG succeeded because the specification was deliberately written with a close correspondence to the implementation, via direct transcription.
The implementation under test was a self-contained function with defined inputs and output, of approximately one thousand lines of code. 
By design of the OT algorithm, the order of clients exchanging operations they performed locally does not change the transformed version of these operations or the resulting array.
As a result of this symmetry, model-checking the specification is feasible, and the implementation can be completely tested by a constrained set of tests that can be generated, compiled, and run fairly quickly.

MBTC and MBTCG are both promising approaches to ensuring conformance between a specification and its implementation.
In a small system, MBTCG was highly effective and accomplished its goal.
When MBTCG is possible, and when the state space is not too large, MBTCG is the gold standard of ensuring specification-implementation conformance.
If the state space becomes too large though, or if any individual test case would take too long to run, we must start sampling the space, and MBTC can be effective.
MBTC is also a good approach when the implementation's behavior is non-deterministic and cannot be forced to execute specific test sequences.
For a very large system, however, either technique may require great effort per specification.

Future work should make both model-based testing approaches more practical.
TLC performance was an obstacle, and any improvements would help.
Better research and tooling to post-process partial states and merge them together under certain assumptions could make the post-processing safer and easier.
More research should determine best practices for writing specifications intended to be used in model-based testing.
It should also determine best practices for logging trace events and generating test-cases from a model.
Developing tooling for whole-process snapshotting could have greatly simplified MBTC trace logging, since we could have used the snapshots to create trace events.
Research into record-and-replay debuggers \cite{OCallahan17RRDebugger} might be applicable to this area.
Making MBTC and MBTCG first-class features in TLC could have saved us a great deal of time.
The state space representation in TLC could be improved to be more easily parsed, which would have eased MBTCG.
Trace-checking could be built in where users only need to provide a trace and a specification and TLC efficiently checks the traces for them.

Going forward, we plan to continue using TLA\textsuperscript{+} to give us confidence and increase our own understanding in our complex protocols.
MongoDB Realm Sync is actively using and improving its use of MBTCG.
More teams and projects throughout MongoDB are starting to use TLA\textsuperscript{+} to gain similar benefits in their own development.
These projects may find interest in ensuring specification-implementation conformance, and this research and any further research will help us determine when the investment is worth it, and what model-based testing method we should use.


\section{Acknowledgments}
\label{sec:acknowledgments}
Many experts have advised us, proposed workarounds for obstacles, reviewed our code, and reviewed drafts of this paper, including 
David Bradford,
Mark Callaghan,
Mike O'Brien,
Siyuan Zhou,
Will Schultz,
Michael Cahill,
Simon Ulsnes,
Tess Avitabile,
Henrik Edin,
Andy Schwerin,
and Ron Pressler.
We are embarrassingly indebted to Markus Alexander Kuppe for answering our questions about TLC within minutes, and implementing features we requested within hours.

\bibliographystyle{abbrv-original-case}
\bibliography{eXtreme-Modelling-in-Practice}  
\end{document}